\documentclass[12pt]{article}

\usepackage{cite}
\usepackage{graphicx}
\usepackage{pict2e}

\author{Yu.~M.~Zinoviev
       \thanks{E-mail address: Yurii.Zinoviev@ihep.ru} \\[0.5cm]
        {\it Institute for High Energy Physics} \\
        {\it of National Research Center "Kurchatov Institute"} \\
        {\it Protvino, Moscow Region, 142280, Russia}}

\title{On massive higher spins and gravity. I. Spin 5/2}

\date{}

\begin{document}

\maketitle

\begin{abstract}
In this paper, we continue our investigation of gravitational
interactions for massive higher spins extending our previous work
on massive spin 3/2 and spin 2 to massive spin 5/2, including
partially massless and massless limits. We use the gauge invariant
frame-like description for massive fields, both for 
general analysis of possible vertices and for constructing 
the minimal vertex (i.e. vertex containing standard minimal
interactions and non-minimal interactions with a minimum number of
derivatives). In particular, we show that there is a
special point $m^2 = 4\Lambda$, which corresponds to the boundary of a
unitary allowed region in $dS_4$, where minimal interactions 
disappear, leaving only non-minimal ones. 
\end{abstract}

\thispagestyle{empty}
\newpage
\setcounter{page}{1}

\section{Introduction}

In this work, we continue our investigation of gravitational
interactions for massive higher spin fields \cite{KhZ21,KhZ21a}. This
is an old topic, but recently there has been a revival of interest
connected with the black holes and gravitational waves 
\cite{CJP21,CP22,CCJOPS22,CCJOPS23,ST23}. We use the frame-like gauge
invariant formalism \cite{Zin08b,PV10,KhZ19} to describe of
massive higher spins. One of the specific features of this 
formalism is the emergence of so-called extra fields, which do not
enter the free Lagrangian but are necessary to construct a
complete set of gauge invariant objects and higher derivative
interactions. As a result, direct constructive approach does not
apply, and we use instead the Fradkin-Vasiliev formalism
\cite{FV87,FV87a,Vas11,BPS12,Zin24a}. One more technical
problem is connected with the existence of Stueckelberg fields, which
are necessary for a gauge invariant description of the massive
fields (both in metric-like and frame-like formalism). Due to their
non-homogeneous transformations, field redefinitions involving these
fields can drastically change the properties of the vertices
\cite{BDGT18,Zin24a}. In particular, for any cubic vertex with one
massless and two massive fields (as in gravitational interactions),
there are always enough field redefinitions to bring it into an
abelian form. Leaving aside questions of physical equivalence 
between models connected by such (higher derivative) field
redefinitions, we use this fact to classify possible vertices. On
the other hand, in order to construct a minimal vertex (i.e.
containing a standard minimal gravitational interaction), we use a
fairly common down-up approach. We start with the standard
substitution rules and then in order to make the vertex completely
gauge invariant, we try to find appropriate non-minimal interactions
with a minimum number of derivatives. In this case,  experience gained
from studying interactions of massless and partially massless fields
greatly simplifies the task.

It has been known for a long time that the spin 5/2 is the first to
require non-minimal gravitational interactions
\cite{AD79,Por93,CPD94} (see also \cite{Sor04}). The cubic interaction
of massless spin 2 and massive, partially massless or massless spin
5/2 was investigated in \cite{Met06a}. There were two limitations in
this work. First, both fields are considered to be on-shell, and
any possible corrections to gauge transformations are not
considered. Second, the number of derivatives is restricted to be
less than three. Our goal here is to expand these results in both
directions.

The paper is structured as follows. In Section 2 we consider e/m
interactions for massive spin 3/2. As is well-known, this is the first
spin where non-minimal e/m interactions are required \cite{FPT92}, and
we use this simple example as an illustration for our general
approach. Section 3 contains all the necessary kinematic information
for a gauge invariant description of a massive spin 5/2 (including
partially massless and massless limits) in a frame-like multispinor
formalism \cite{KhZ19}. In Section 4, we discuss interactions for
massless spin 5/2. In this case there are no any ambiguities
associated with field redefinitions, so using the Fradkin-Vasiliev
formalism we can directly obtain the vertex, which includes both
minimal interactions as well as non-minimal interactions with no more
than two derivatives. Section 5 is devoted to the partially massless
case. An analysis of all possible field redefinitions showed that
there is just one non-abelian vertex (i.e. that cannot be made abelian
through field redefinitions). To construct this vertex, we use the
Skvortsov-Vasiliev version \cite{SV06} for a frame-like description of
the partially massless spin 5/2. The resulting vertex also contains
both minimal interactions and non-minimal interacttions whith no more
than two derivatives. Finally, in Section 6, we consider all possible
interactions for massive spin 5/2, including minimal one. 

\noindent
{\bf Notations and conventions} In the frame-like multispinor
formalism we use \cite{KhZ19}, all objects are forms (fields are
one-forms or zero-forms, gauge invariant curvatures are two-forms or
one-forms, while all the terms in Lagrangians are four-forms)
with some number of completely symmetric dotted and un-dotted spinor
indices $\alpha,\dot\alpha = 1,2$. The coordinate-free description of
(anti-) de Sitter space is achieved using the background frame
$e^{\alpha\dot\alpha}$ and the covariant external derivative $D$
(which contains a background Lorentz connection) which satisfy
\begin{equation}
D \wedge e^{\alpha\dot\alpha} = 0, \qquad
D \wedge D \zeta^\alpha = 2\Lambda E^\alpha{}_\beta \zeta^\beta.
\end{equation}
Here basic two-forms $E^{\alpha(2)}$, $E^{\dot\alpha(2)}$ are defined
as follows
\begin{equation}
e^{\alpha\dot\alpha} \wedge e^{\beta\dot\beta} =
\epsilon^{\alpha\beta} E^{\dot\alpha\dot\beta}
+ e^{\dot\alpha\dot\beta} E^{\alpha\beta}.
\end{equation}
A complete basis of forms also contains three-form
$E^{\alpha\dot\alpha}$ and four-form $E$:
\begin{equation}
E^{\alpha(2)} \wedge e^{\beta\dot\alpha} =
\epsilon^{\alpha\beta} E^{\alpha\dot\alpha}, \qquad
E^{\alpha\dot\alpha} \wedge e^{\beta\dot\beta}
= \epsilon^{\alpha\beta} \epsilon^{\dot\alpha\dot\beta} E.
\end{equation}
In what follows all wedge product signs are omitted.

\section{Spin 3/2 and e/m field}

In this section we illustrate our approach  using the simple example
of e/m interactions for massive spin 3/2 (see e.g.
\cite{FPT92,DPW00,Zin06}). Recall that for e/m interactions the spin
3/2 is the first one that requires non-minimal corrections.

\subsection{Kinematics}

{\bf Spin 3/2} In this case we use one-form $\Phi^\alpha + h.c.$ and
zero-form  $\phi^\alpha + h.c.$. The free Lagrangian looks like:
\begin{eqnarray}
{\cal L}_0 &=& - \Phi_\alpha e^\alpha{}_{\dot\alpha} D
\Phi^{\dot\alpha} - \phi_\alpha E^\alpha{}_{\dot\alpha} D
\phi^{\dot\alpha} \nonumber \\
 && - M \Phi_\alpha E^\alpha{}_\beta \Phi^\beta 
 + a_0 \Phi_\alpha E^\alpha{}_{\dot\alpha} 
 + M E \phi_\alpha \phi^\alpha + h.c. 
\end{eqnarray}
It is invariant under the following gauge transformations
\begin{equation}
\delta \Phi^\alpha = D \rho^\alpha + M 
e^\alpha{}_{\dot\alpha} \rho^{\dot\alpha}, \qquad
\delta \phi^\alpha = a_0 \rho^\alpha,
\end{equation}
where
$$
M^2 = m^2 + \lambda^2, \qquad a_0{}^2 = 6m^2.
$$
Corresponding gauge invariant curvatures
\begin{eqnarray}
{\cal F}^\alpha &=& D \Phi^\alpha + M e^\alpha{}_{\dot\alpha}
\Phi^{\dot\alpha} - \frac{a_0}{3} E^\alpha{}_\beta \phi^\beta,
\nonumber \\
{\cal C}^\alpha &=& D \phi^\alpha - a_0 \Phi^\alpha + M 
e^\alpha{}_{\dot\alpha} \phi^{\dot\alpha}, 
\end{eqnarray}
satisfy the following differential identities
\begin{eqnarray}
D {\cal F}^\alpha &=& - M e^\alpha{}_{\dot\alpha} 
{\cal F}^{\dot\alpha} - \frac{a_0}{3} E^\alpha{}_\beta {\cal C}^\beta,
\nonumber \\
D {\cal C}^\alpha &=& - a_0 {\cal F}^\alpha - M 
e^\alpha{}_{\dot\alpha} {\cal C}^{\dot\alpha}.
\end{eqnarray}
Variation of the free Lagrangian under the arbitrary field variations
can be written as
\begin{equation}
\delta {\cal L}_0 = - {\cal F}_{\dot\alpha} e_\alpha{}^{\dot\alpha}
\delta \Phi^\alpha - {\cal C}_{\dot\alpha} E_\alpha{}^{\dot\alpha}
\delta \phi^\alpha + h.c.
\end{equation}
For the second spin 3/2 we use notation $\Psi^\alpha$, $\psi^\alpha$
and $\zeta^\alpha$. \\
{\bf Spin 1} We use a one-form $A$ and zero-forms $B^{\alpha(2)}$,
$B^{\dot\alpha(2)}$. The free Lagrangian
\begin{equation}
{\cal L}_0 = 4E B_{\alpha(2)} B^{\alpha(2)} + 2 E_{\alpha(2)}
B^{\alpha(2)} D A + h.c.
\end{equation}
A gauge invariant curvature is simply
\begin{equation}
{\cal B}^{\alpha(2)} = D B^{\alpha(2)}.
\end{equation}
By analogy with torsion zero condition in gravity, we work partially
on-shell, i.e. on the auxiliary fields equations so that
\begin{equation}
E_{\alpha(2)} {\cal B}^{\alpha(2)} + E_{\dot\alpha(2)}
{\cal B}^{\dot\alpha(2)} \approx 0, \qquad
DA = 2 E_{\alpha(2)} B^{\alpha(2)} - 2 E_{\dot\alpha(2)}
B^{\dot\alpha(2)}.
\end{equation}
Any variation of the physical field $A$ produces
\begin{equation}
\delta {\cal L}_0 = - 2 E_{\alpha(2)} {\cal B}^{\alpha(2)} \delta A +
h.c. 
\end{equation}

\subsection{Minimal vertex}

To obtain a minimal vertex we follow a constructive approach
restricting ourselves with the minimal number of derivatives. We
begin with the following ansatz for supertransformations
\begin{eqnarray}
\delta \Phi^\alpha &=& q A \zeta^\alpha + a_1 B^{\alpha\beta}
e_{\beta\dot\alpha} \zeta^{\dot\alpha}, \qquad
\delta \phi^\alpha = a_2 B^{\alpha\beta} \zeta_\beta, \nonumber \\
\delta A &=& b_1 \Phi^\alpha \zeta_\alpha + b_2 e_{\alpha\dot\alpha}
\phi^{\dot\alpha} \zeta^\alpha + h.c. 
\end{eqnarray}
Bosonic variations give
\begin{equation}
\delta {\cal L}_B = - 2b_1 \Phi_\alpha E_{\dot\alpha(2)} 
{\cal B}^{\dot\alpha(2)} \zeta^\alpha - 4b_2 \phi_{\dot\alpha}
E_\alpha{}^{\dot\alpha} {\cal B}^{\alpha\beta} \zeta_\beta + h.c.
\end{equation}
while fermionic variations look like
\begin{equation}
\delta {\cal L}_F = - q {\cal F}_{\dot\alpha} e_\alpha{}^{\dot\alpha}
A \zeta^\alpha + a_1 {\cal F}_\alpha e^\alpha{}_{\dot\alpha}
B^{\dot\alpha\dot\beta} e_{\beta\dot\beta} \zeta^\beta
- a_2 {\cal C}_{\dot\alpha} E_\alpha{}^{\dot\alpha} B^{\alpha\beta}
\zeta_\beta + h.c. 
\end{equation}
Using explicit expressions for curvatures and integrating by parts
this variations can be rewritten in the on-shell equivalent form
\begin{eqnarray}
\delta {\cal L}_F &=& - a_1 \Phi_\alpha E_{\dot\alpha(2)} 
{\cal B}^{\dot\alpha(2)} \zeta^\alpha - a_2 \phi_{\dot\alpha}
E_\alpha{}^{\dot\alpha} {\cal B}^{\alpha\beta} \zeta_\beta \nonumber 
\\
 && + [ - q \Phi_{\dot\alpha} e_\alpha{}^{\dot\alpha} A - a_1
\Phi_\alpha E_{\dot\alpha(2)} B^{\dot\alpha(2)} + a_2
\phi_{\dot\alpha} E_\beta{}^{\dot\alpha} B^\beta{}_\alpha]
 [ D \zeta^\alpha + M e^\alpha{}_{\dot\beta} \zeta^{\dot\beta}]
\nonumber \\
 && + [ 4Ma_1 + a_0a_2 - 8q] \Phi_{\dot\alpha} E_\alpha{}^{\dot\alpha}
B^{\alpha\beta} \zeta_\beta - a_0q \phi_{\dot\alpha} 
E^{\alpha\dot\alpha} A \zeta_\alpha + 2Ma_2 E \phi_\alpha
B^{\alpha\beta} \zeta_\beta.
\end{eqnarray}
Terms in the first line compensate bosonic variations at
\begin{equation}
b_1 = - \frac{a_1}{2}, \qquad b_2 = - \frac{a_2}{4}.
\end{equation}
The second line contains supercurrent and can be compensated by the
corresponding interactions with the field $\Psi_\alpha$, while to
compensate terms in the last line we have to put
\begin{equation}
8q = 4Ma_1 + a_0a_2 
\end{equation}
and add interactions with goldstino $\psi_\alpha$. We obtain a cubic
vertex:
\begin{eqnarray}
{\cal L}_1 &=& - q \Phi_{\dot\alpha} e^{\alpha\dot\alpha} A
\Psi_\alpha + q \phi_{\dot\alpha} E^{\alpha\dot\alpha} A \psi_\alpha
\nonumber \\
 && + a_1 \Phi_\alpha E_{\dot\alpha(2)} B^{\dot\alpha(2)} \Psi^\alpha
+ a_2 ( \phi_{\dot\alpha} E_\alpha{}^{\dot\alpha} B^{\alpha\beta}
\Psi_\beta - \psi_{\dot\alpha} E_\alpha{}^{\dot\alpha} B^{\alpha\beta}
\Phi_\beta) - \frac{2Ma_2}{a_0} E \phi_\alpha B^{\alpha\beta}
\psi_\beta 
\end{eqnarray}
where apart from the standard minimal terms in the first line we have
non-minimal terms that are necessary for the vertex to be invariant
under the local supertransformations $\rho^\alpha$, $\zeta^\alpha$.

\subsection{General analysis}

Here we relax the restriction on the number of derivatives and allow
all possible field redefinitions\footnote{Similar analysis in the
BV-formalism can be found in \cite{BLT24}}. We have two massive
fields and one massless field, and any such vertex can be transformed
into an abelian one \cite{Zin24a}. Let us consider the most general
abelian vertex (we do not assume that the two masses are equal):
\begin{equation}
{\cal L}_1 = g_1 {\cal F}^\alpha \tilde{\cal C}_\alpha A + g_2 
{\cal C}^\alpha \tilde{\cal F}_\alpha A + g_3 {\cal C}_\alpha
e^{\alpha\dot\alpha} \tilde{\cal C}_{\dot\alpha} A + h.c. 
\end{equation}
Calculating its variation we obtain
\begin{eqnarray}
\delta {\cal L}_1 &=& - (\tilde{a}_0g_1 + a_0g_2) {\cal F}^\alpha
\tilde{\cal F}_\alpha \xi \nonumber \\
 && + [(M+\tilde{M})g_1 + a_0g_3] e_\alpha{}^{\dot\alpha}
{\cal F}^\alpha \tilde{\cal C}_{\dot\alpha} \xi
+ [- (M+\tilde{M})g_2 + \tilde{a}_0g_3] {\cal C}^\alpha
e_\alpha{}^{\dot\alpha} \tilde{\cal F}_{\dot\alpha} \xi \nonumber \\
 && + [\frac{a_0}{3}g_1 + \frac{\tilde{a}_0}{3}g_2 + 2
(M-\tilde{M})g_3] E^{\alpha\beta} {\cal C}_\alpha
\tilde{\cal C}_\beta \xi + h.c. 
\end{eqnarray}
Terms in the second line can be compensated by corrections
$$
\delta \Phi^\alpha \sim \tilde{\cal C}^\alpha \xi, \qquad 
\delta \Psi^\alpha \sim {\cal C}^\alpha \xi,
$$
so we have to put
$$
\tilde{a}_0g_1 + a_0g_2 = 0, \qquad
a_0g_1 + \tilde{a}_0g_2 = - 6 (M-\tilde{M})g_3.
$$
If we set
$$
g_2 = - \frac{\tilde{a}_0}{a_0}g_1,
$$
we obtain
$$
(M-\tilde{M})g_3 = - \frac{M^2 - \tilde{M}^2}{a_0}g_1.
$$
If two masses are not equal we have only one solution. Moreover it
can be shown that such vertex is equivalent to the trivially gauge
invariant one
\begin{equation}
{\cal L}_1 \sim {\cal C}^\alpha \tilde{\cal C}_\alpha D A + h.c.
\end{equation}
But if two masses are equal we have two independent solutions:
\begin{equation}
{\cal L}_1 = g_1 [ {\cal F}^\alpha \tilde{\cal C}_\alpha A - 
{\cal C}^\alpha \tilde{\cal F}_\alpha A] + g_3 {\cal C}_\alpha
e^{\alpha\dot\alpha} \tilde{\cal C}_{\dot\alpha} A + h.c. 
\end{equation}
In the unitary gauge (that is when the Stueckelberg zero-forms
$\phi^\alpha$ and $\psi^\alpha$ are set to zero) we have
\begin{equation}
{\cal L}_1 = a_0g_1 \Phi^\alpha \Psi_\alpha D A + a_0{}^2g_3
\Phi_\alpha e^{\alpha\dot\alpha} \Psi_{\dot\alpha} A + h.c. 
\end{equation}
Thus some particular combination must correspond to the minimal
vertex. 

\section{Kinematics}

In this section we provide all necessary kinematic information on
the massive, partially massless and massless spin 5/2 as well as on
massless spin 2.

\subsection{Massive spin 5/2}

In $d=4$ a massive spin 5/2 has six helicities $(\pm 5/2, \pm 3/2, \pm
1/2)$ so the gauge invariant frame-like formalism \cite{KhZ19}
requires one-forms $\Phi^{\alpha(2)\dot\alpha} + h.c.$, 
$\Phi^\alpha + h.c.$ and zero-form $\phi^\alpha + h.c.$. The free
Lagrangian (four-form in our formalism) has the form
\begin{eqnarray} 
{\cal L}_0 &=& - D \Phi_{\alpha\beta\dot\alpha} e^\beta{}_{\dot\beta}
\Phi^{\alpha\dot\alpha\dot\beta} + D \Phi_\alpha 
e^\alpha{}_{\dot\alpha} \Phi^{\dot\alpha}- 3a_0 D \phi_\alpha
E^\alpha{}_{\dot\alpha} \phi^{\dot\alpha} \nonumber \\
 && + \frac{M}{2} [ 3 \Phi_{\alpha\beta\dot\alpha} E^\beta{}_\gamma
\Phi^{\alpha\gamma\dot\alpha} - \Phi_{\alpha(2)\dot\alpha}
E^{\dot\alpha}{}_{\dot\beta} \Phi^{\alpha(2)\dot\beta}] 
 + 2\tilde{m} \Phi_{\alpha(2)\dot\alpha} E^{\alpha(2)}
\Phi^{\dot\alpha} \nonumber \\
 && - 3M \Phi_\alpha E^\alpha{}_\beta \Phi^\beta + 6a_0
\Phi_\alpha E^\alpha{}_{\dot\alpha} \phi^{\dot\alpha}
+ 9Ma_0 E \phi_\alpha \phi^\alpha + h.c. \label{lag_m}
\end{eqnarray}
where
\begin{equation}
M^2 = m^2 - 4\Lambda, \qquad
\tilde{m}^2 = \frac{5}{4}m^2, \qquad
a_0 = \frac{16}{3}(m^2 - 3\Lambda).
\end{equation}
This Lagrangian is invariant under the following local gauge
transformations
\begin{eqnarray}
\delta \Phi^{\alpha(2)\dot\alpha} &=& D \rho^{\alpha(2)\dot\alpha}
+ e_\beta{}^{\dot\alpha} \rho^{\alpha(2)\beta} + \frac{M}{2}
e^\alpha{}_{\dot\beta} \rho^{\alpha\dot\alpha\dot\beta} + 
\frac{\tilde{m}}{3} e^{\alpha\dot\alpha} \rho^\alpha, \nonumber \\
\delta \Phi^\alpha &=& D \rho^\alpha + \tilde{m} e_{\beta\dot\alpha}
\rho^{\alpha\beta\dot\alpha} + \frac{3M}{2} e^\alpha{}_{\dot\alpha}
\rho^{\dot\alpha}, \label{gauge_1} \\
\delta \phi^\alpha &=& \rho^\alpha. \nonumber
\end{eqnarray}
To construct a complete set of gauge invariant objects (curvatures)
we need extra fields: one-form $\Phi^{\alpha(3)} + h.c.$ and 
zero-forms $\phi^{\alpha(3)} + h.c.$, $\phi^{\alpha(2)\dot\alpha} +
h.c.$ with gauge transformations
\begin{equation}
\delta \Phi^{\alpha(3)} = D \rho^{\alpha(3)} + \frac{a_0}{16} 
e^\alpha{}_{\dot\alpha} \rho^{\alpha(2)\dot\alpha}, \qquad
\delta \phi^{\alpha(3)} = \rho^{\alpha(3)}, \qquad
\delta \phi^{\alpha(2)\dot\alpha} = \rho^{\alpha(2)\dot\alpha}.
\label{gauge_2}
\end{equation}
Thus the complete set of fields looks like:
$$
\begin{array}{|c|c|c|c|} \hline
one-forms & \Phi^{\alpha(3)} & \Phi^{\alpha(2)\dot\alpha} &
\Phi^\alpha \\ \hline
zero-forms & \phi^{\alpha(3)} & \phi^{\alpha(2)\dot\alpha} &
\phi^\alpha \\ \hline \end{array} + h.c. 
$$
and we have a one-to-one correspondence between one-forms and
zero-forms.

We obtain gauge invariant two-forms 
\begin{eqnarray}
{\cal F}^{\alpha(3)} &=& D \Phi^{\alpha(3)} + \frac{a_0}{16} 
e^\alpha{}_{\dot\alpha} \Phi^{\alpha(2)\dot\alpha} - \frac{2m^2}{3}
E^\alpha{}_\beta \phi^{\alpha(2)\beta} - \frac{\tilde{m}a_0}{12}
E^{\alpha(2)} \phi^\alpha, \nonumber \\
{\cal F}^{\alpha(2)\dot\alpha} &=& D \Phi^{\alpha(2)\dot\alpha} + 
e_\beta{}^{\dot\alpha} \Phi^{\alpha(2)\beta} + \frac{M}{2} 
e^\alpha{}_{\dot\beta} \Phi^{\alpha\dot\alpha\dot\beta} +
\frac{\tilde{m}}{3} e^{\alpha\dot\alpha} \Phi^\alpha, \label{cur_1} \\
{\cal F}^\alpha &=& D \Phi^\alpha + \tilde{m} e_{\beta\dot\alpha}
\Phi^{\alpha\beta\dot\alpha} + \frac{3M}{2} e^\alpha{}_{\dot\alpha}
\Phi^{\dot\alpha} - 2\tilde{m} E_{\beta(2)} \phi^{\alpha\beta(2)} -
a_0 E^\alpha{}_\beta \phi^\beta \nonumber
\end{eqnarray}
and gauge invariant one-forms
\begin{eqnarray}
{\cal C}^{\alpha(3)} &=& D \phi^{\alpha(3)} - \Phi^{\alpha(3)} 
+ \frac{a_0}{16} e^\alpha{}_{\dot\alpha} \phi^{\alpha(2)\dot\alpha},
\nonumber \\
{\cal C}^{\alpha(2)\dot\alpha} &=& D \phi^{\alpha(2)\dot\alpha} 
- \Phi^{\alpha(2)\dot\alpha} + e_\beta{}^{\dot\alpha}
\phi^{\alpha(2)\beta} + \frac{M}{2} e^\alpha{}_{\dot\beta}
\phi^{\alpha\dot\alpha\dot\beta} + \frac{\tilde{m}}{3}
e^{\alpha\dot\alpha} \phi^\alpha, \label{cur_2} \\
{\cal C}^\alpha &=& D \phi^\alpha - \Phi^\alpha + \tilde{m} 
e_{\beta\dot\alpha} \phi^{\alpha\beta\dot\alpha} + \frac{3M}{2}
e^\alpha{}_{\dot\alpha} \phi^{\dot\alpha}. \nonumber
\end{eqnarray}
The extra fields do not enter the free Lagrangian so in what follows
we do not use a term "on equations of motions" and use a term 
"on-shell". In this particular case we have
\begin{equation}
{\cal F}^{\alpha(2)\dot\alpha} \approx 0, \qquad
{\cal F}^\alpha \approx 0, \qquad
{\cal C}^\alpha \approx 0,
\end{equation}
while
\begin{equation}
{\cal F}^{\alpha(3)} \approx E_{\beta(2)} Y^{\alpha(3)\beta(2)},
\qquad {\cal C}^{\alpha(3)} \approx e_{\beta\dot\beta}
Y^{\alpha(3)\beta\dot\beta}, \qquad {\cal C}^{\alpha(2)\dot\alpha}
\approx e_{\beta\dot\beta} Y^{\alpha(2)\beta\dot\alpha\dot\beta}.
\end{equation}
Here $Y^{\alpha(5)}$, $Y^{\alpha(4)\dot\alpha}$ and 
$Y^{\alpha(3)\dot\alpha(2)}$ are just the first representatives of the
infinite set of gauge invariant zero-forms satisfying the so-called
unfolded equations \cite{KhZ19}. The three curvatures that are 
non-zero on-shell satisfy the following differential and algebraic
identities:
\begin{eqnarray}
D {\cal F}^{\alpha(3)} &\approx& - \frac{2m^2}{3} E^\alpha{}_\beta
{\cal C}^{\alpha(2)\beta}, \nonumber \\
D {\cal C}^{\alpha(3)} &\approx& - {\cal F}^{\alpha(3)} 
- \frac{a_0}{16} e^\alpha{}_{\dot\alpha} 
{\cal C}^{\alpha(2)\dot\alpha}, \\
D {\cal C}^{\alpha(2)\dot\alpha} &\approx& - e_\beta{}^{\dot\alpha}
{\cal C}^{\alpha(2)\beta} - \frac{M}{2} e^\alpha{}_{\dot\beta}
{\cal C}^{\alpha\dot\alpha\dot\beta}, \nonumber
\end{eqnarray}
\begin{equation}
e_\beta{}^{\dot\alpha} {\cal F}^{\alpha(2)\beta} \approx 0, \qquad
E_{\beta(2)} {\cal C}^{\alpha\beta(2)} \approx 0, \qquad
e_{\beta\dot\alpha} {\cal C}^{\alpha\beta\dot\alpha} \approx 0.
\end{equation}

Variations of the free Lagrangian under the arbitrary variations of
the physical fields can be calculated as follows:
\begin{equation}
\delta {\cal L}_0 = 2 {\cal F}_{\alpha\dot\alpha\dot\beta}
e_\beta{}^{\dot\beta} \delta \Phi^{\alpha\beta\dot\alpha} 
- 2 {\cal F}_{\dot\alpha} e_\alpha{}^{\dot\alpha} \delta \Phi^\alpha 
- 6a_0 {\cal C}_{\dot\alpha} E_\alpha{}^{\dot\alpha} \delta
\phi^\alpha + h.c. 
\end{equation}

\subsection{Partially massless spin 5/2}

Recall that in $dS_4$ there exists an unitary forbidden region 
$m^2 < 4\Lambda$ for the massive spin 5/2. Inside this region at 
$m^2 = 3\Lambda$ one finds a so-called partially massless spin 5/2.
\begin{figure}[htb]
\setlength{\unitlength}{0.8mm}
\begin{center}
\begin{picture}(80,60)
\put(10,20){\vector(1,0){60}}
\put(10,0){\vector(0,1){50}}
\put(0,45){\makebox(10,10)[]{$\Lambda$}}
\put(70,15){\makebox(10,10)[]{$m^2$}}

\put(10,20){\line(2,1){60}}
\put(10,20){\line(1,1){30}}
\put(38,50){\makebox(10,10)[]{$m^2=3\Lambda$}}
\put(68,50){\makebox(10,10)[]{$m^2=4\Lambda$}}
\end{picture}
\caption{Unitary forbidden region $m^2 < 4\Lambda$ for massive spin
5/2}
\end{center}
\end{figure}
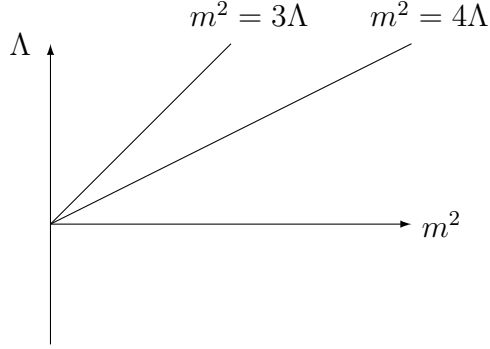
In this case we have $a_0 = 0$ and so the spin 1/2 component 
$\phi^\alpha$ decouples leaving us with the Lagrangian
\begin{eqnarray}
{\cal L}_0 &=& - D \Phi_{\alpha\beta\dot\alpha} e^\beta{}_{\dot\beta}
\Phi^{\alpha\dot\alpha\dot\beta} + D \Phi_\alpha 
e^\alpha{}_{\dot\alpha} \Phi^{\dot\alpha} \nonumber \\
 && + \frac{M}{2} [ 3 \Phi_{\alpha\beta\dot\alpha} E^\beta{}_\gamma
\Phi^{\alpha\gamma\dot\alpha} - \Phi_{\alpha(2)\dot\alpha}
E^{\dot\alpha}{}_{\dot\beta} \Phi^{\alpha(2)\dot\beta}] \nonumber \\
 && + 2\tilde{m} \Phi_{\alpha(2)\dot\alpha} E^{\alpha(2)}
\Phi^{\dot\alpha} - 3M \Phi_\alpha E^\alpha{}_\beta \Phi^\beta, 
\end{eqnarray}
which is still invariant under the gauge transformations
\begin{eqnarray}
\delta \Phi^{\alpha(2)\dot\alpha} &=& D \rho^{\alpha(2)\dot\alpha}
 + e_\beta{}^{\dot\alpha} \rho^{\alpha(2)\beta}
 + \frac{M}{2} e^\alpha{}_{\dot\beta} \rho^{\alpha\dot\alpha\dot\beta}
+ \frac{\tilde{m}}{3} e^{\alpha\dot\alpha} \rho^\alpha, \nonumber \\
\delta \Phi^\alpha &=& D \rho^\alpha + \tilde{m} e_{\beta\dot\alpha}
\rho^{\alpha\beta\dot\alpha} + \frac{3M}{2} e^\alpha{}_{\dot\alpha}
\rho^{\dot\alpha}. 
\end{eqnarray}
The remaining gauge invariant curvatures look like:
\begin{eqnarray}
{\cal F}^{\alpha(3)} &=& D \Phi^{\alpha(3)} - \frac{2m^2}{3}
E^\alpha{}_\beta \phi^{\alpha(2)\beta}, \nonumber \\
{\cal F}^{\alpha(2)\dot\alpha} &=& D \Phi^{\alpha(2)\dot\alpha} + 
e_\beta{}^{\dot\alpha} \Phi^{\alpha(2)\beta} + 
\frac{M}{2} e^\alpha{}_{\dot\beta} \Phi^{\alpha\dot\alpha\dot\beta} +
\frac{\tilde{m}}{3} e^{\alpha\dot\alpha} \Phi^\alpha, \nonumber \\
{\cal F}^\alpha &=& D \Phi^\alpha + \tilde{m} e_{\beta\dot\alpha}
\Phi^{\alpha\beta\dot\alpha} + \frac{3M}{2} e^\alpha{}_{\dot\alpha}
\Phi^{\dot\alpha} - 2\tilde{m} E_{\beta(2)} \phi^{\alpha\beta(2)}, \\
{\cal C}^{\alpha(3)} &=& D \phi^{\alpha(3)} - \Phi^{\alpha(3)}.
\nonumber
\end{eqnarray}
In \cite{SV06} Skvortsov and Vasiliev proposed a very compact and
convenient description for bosonic partially massless fields. In 
\cite{KhZ19} we showed that such a description corresponds to the
partial gauge fixing of the general formalism, which arises from the
massive one for special values of mass and extend this procedure
to fermionic fields. For partially massless spin 5/2, this
corresponds to setting $\phi^{\alpha(3)} = 0$ and solving its
equation (see Appendix B of \cite{Zin24a} for more details).
This leaves us with one-forms $\Phi^{\alpha(2)\dot\alpha} + h.c.$,
$\Phi^\alpha + h.c.$ only and with the corresponding two-forms
\begin{eqnarray}
{\cal F}^{\alpha(2)\dot\alpha} &=& D \Phi^{\alpha(2)\dot\alpha} + 
\frac{M}{2} e^\alpha{}_{\dot\beta} \Phi^{\alpha\dot\alpha\dot\beta} +
\frac{\tilde{m}}{3} e^{\alpha\dot\alpha} \Phi^\alpha, \nonumber \\
{\cal F}^\alpha &=& D \Phi^\alpha + \tilde{m} e_{\beta\dot\alpha}
\Phi^{\alpha\beta\dot\alpha} + \frac{3M}{2} e^\alpha{}_{\dot\alpha}
\Phi^{\dot\alpha}.  
\end{eqnarray}
Now the free Lagrangian can be rewritten in the explicitly gauge
invariant form: 
\begin{equation}
{\cal L}_0 = - \frac{1}{2M} {\cal F}_{\alpha(2)\dot\alpha}
{\cal F}^{\alpha(2)\dot\alpha} + \frac{1}{3M} {\cal F}_\alpha
{\cal F}^\alpha + h.c. 
\end{equation}

\subsection{Massless spin 5/2}

In this case we have
\begin{equation}
m = 0 \quad \Rightarrow \quad M^2 = 4\lambda^2, \qquad 
a_0 = 16\lambda^2.
\end{equation}
The free Lagrangian is simply
\begin{equation}
{\cal L}_0 = - D \Phi_{\alpha\beta\dot\alpha} e^\beta{}_{\dot\beta}
\Phi^{\alpha\dot\alpha\dot\beta} + \lambda [ 3
\Phi_{\alpha\beta\dot\alpha} E^\beta{}_\gamma
\Phi^{\alpha\gamma\dot\alpha} - \Phi_{\alpha(2)\dot\alpha}
E^{\dot\alpha}{}_{\dot\beta} \Phi^{\alpha(2)\dot\beta}] + h.c.
\end{equation}
It is invariant under the gauge transformations
\begin{equation}
\delta \Phi^{\alpha(2)\dot\alpha} = D \rho^{\alpha(2)\dot\alpha}
+ e_\beta{}^{\dot\alpha} \rho^{\alpha(2)\beta} + \lambda
e^\alpha{}_{\dot\beta} \rho^{\alpha\dot\alpha\dot\beta}.
\end{equation}
To construct a complete set of the gauge invariant curvatures we still
need  an extra one-form $\Phi^{\alpha(3)} + h.c.$ with gauge
transformations
\begin{equation}
\delta \Phi^{\alpha(3)} = D \rho^{\alpha(3)} + \lambda^2
e^\alpha{}_{\dot\alpha} \rho^{\alpha(2)\dot\alpha}.
\end{equation}
The gauge invariant curvatures look like
\begin{eqnarray}
{\cal F}^{\alpha(3)}  &=& D \Phi^{\alpha(3)} + \lambda^2 
e^\alpha{}_{\dot\alpha} \Phi^{\alpha(2)\dot\alpha}, \nonumber \\
{\cal F}^{\alpha(2)\dot\alpha} &=& D \Phi^{\alpha(2)\dot\alpha}
+ e_\beta{}^{\dot\alpha} \Phi^{\alpha(2)\beta} + \lambda
e^\alpha{}_{\dot\beta} \Phi^{\alpha\dot\alpha\dot\beta}.
\end{eqnarray}
The free Lagrangian can be written as follows
\begin{equation}
{\cal L}_0 = - \frac{1}{12\lambda^3} {\cal F}_{\alpha(3)} 
{\cal F}^{\alpha(3)} - \frac{1}{4\lambda} 
{\cal F}_{\alpha(2)\dot\alpha} {\cal F}^{\alpha(2)\dot\alpha} + h.c.
\end{equation}

\subsection{Massless spin 2}

For the frame-like formalism we need one-forms $\omega^{\alpha(2)} +
h.c.$ and $h^{\alpha\dot\alpha}$  with the gauge transformations
\begin{eqnarray}
\delta \omega^{\alpha(2)} &=& D \eta^{\alpha(2)} + \lambda^2 
e^\alpha{}_{\dot\alpha} \xi^{\alpha\dot\alpha}, \nonumber \\
\delta h^{\alpha\dot\alpha} &=& D \xi^{\alpha\dot\alpha} +
e_\beta{}^{\dot\alpha} \eta^{\alpha\beta} + e^\alpha{}_{\dot\beta}
\eta^{\dot\alpha\dot\beta}.
\end{eqnarray}
The corresponding gauge invariant two-forms look like
\begin{eqnarray}
R^{\alpha(2)} &=& D \omega^{\alpha(2)} + \lambda^2
e^\alpha{}_{\dot\alpha} h^{\alpha\dot\alpha}, \nonumber \\
T^{\alpha\dot\alpha} &=& D h^{\alpha\dot\alpha} + 
e_\beta{}^{\dot\alpha} \omega^{\alpha\beta} + e^\alpha{}_{\dot\beta}
\omega^{\dot\alpha\dot\beta}. 
\end{eqnarray}
As it common in the frame-like formalism in gravity and supergravity,
we will use a so-called "torsion zero condition":
\begin{equation}
T^{\alpha\dot\alpha} \approx 0 \quad \Rightarrow \quad
D R^{\alpha(2)} \approx 0, \qquad
e_\beta{}^{\dot\alpha} R^{\alpha\beta} +
e^\alpha{}_{\dot\beta} R^{\dot\alpha\dot\beta} \approx 0.
\end{equation}
The free Lagrangian can be written as:
\begin{equation}
{\cal L}_0 = \frac{i}{4\lambda^2} R_{\alpha(2)} R^{\alpha(2)} + h.c.
\end{equation}
while its variation under the arbitrary variations of the physical
field can be calculated as follows
\begin{equation}
\delta {\cal L}_0 = R_{\alpha\beta} e^\beta{}_{\dot\alpha}
\delta h^{\alpha\dot\alpha} + h.c. 
\end{equation}

\section{Massless case}

In $d=4$ there exist two types of cubic vertices for the massless
fields \cite{Met18a,Met22}. One of these belongs to the so-called
trivially invariant vertices with the number of derivatives
$N_B = s_1 + s_2 + s_3$ for bosonic vertices and $N_F = s_1 + s_2 +
s_3 -1$ for fermionic ones. In \cite{Zin24a} it was shown that if
the so-called triangular inequality  $s_1 \le s_2 + s_3$ holds
(here we assume that $s_1 \ge s_2 \ge s_3$), such vertices can be
expressed using gauge invariant zero-forms. For the specific case 
$(5/2,5/2,2)$, the corresponding vertex can be written in two on-shell
equivalent forms:
\begin{equation}
{\cal L} \sim E W^{\alpha(2)\beta(2)}Y_{\alpha(2)\gamma(3)}
Y_{\beta(2)}{}^{\gamma(3)} \approx W^{\alpha(2)\beta(2)}
{\cal F}_{\alpha(2)\gamma} {\cal F}_{\beta(2)}{}^\gamma,
\end{equation}
where
$$
R^{\alpha(2)} \approx E_{\beta(2)} W^{\alpha(2)\beta(2)}.
$$
For the vertices of the second type the number of derivatives is 
$N_B = s_1 + s_2 - s_3$ for the bosonic case and $N_F = s_1 + s_2 -
s_3 - 1$  for the fermionic one. Moreover, if the so-called strict
triangular inequality $s_2 < s_2 + s_3$ holds, the corresponding
vertex appears to be non-abelian.

To construct a non-abelian vertex $(5/2,5/2,2)$ we use the 
Fradkin-Vasiliev formalism \cite{FV87,FV87a,Vas11,KhZ20a,Zin24a}.
Recall that the first step is to find consistent deformations of all
gauge invariant curvatures. Here, consistency means that the
deformed curvatures $\hat{\cal F} = {\cal F} + \Delta {\cal F}$
transform covariantly $\delta \hat{\cal F} \sim {\cal F}$. For the
spin 5/2 we obtain (we set coupling constant to be 1):
\begin{eqnarray}
\Delta {\cal F}^{\alpha(3)} &=& \omega^\alpha{}_\beta
\Phi^{\alpha(2)\beta} + \lambda^2 h^\alpha{}_{\dot\alpha}
\Phi^{\alpha(2)\dot\alpha}, \nonumber \\
\Delta {\cal F}^{\alpha(2)\dot\alpha} &=& \omega^\alpha{}_\beta
\Phi^{\alpha\beta\dot\alpha} + \omega^{\dot\alpha}{}_{\dot\beta}
\Phi^{\alpha(2)\dot\beta} + h_\beta{}^{\dot\alpha}
\Phi^{\alpha(2)\beta} + \lambda h^\alpha{}_{\dot\beta}
\Phi^{\alpha\dot\alpha\dot\beta}, 
\end{eqnarray}
which corresponds to the standard substitution rules. For the graviton
we obtain:
\begin{eqnarray}
\Delta R^{\alpha(2)} &=& b_1 [ \Phi^{\alpha\beta(2)}
\Phi^\alpha{}_{\beta(2)} + 2\lambda^2 \Phi^{\alpha\beta\dot\alpha}
\Phi^\alpha{}_{\beta\dot\alpha} + \lambda^2 \Phi^{\alpha\dot\alpha(2)}
\Phi^\alpha{}_{\dot\alpha(2)}], \nonumber \\
\Delta T^{\alpha\dot\alpha} &=& b_1 [ 2 \Phi^{\alpha\beta(2)}
\Phi_{\beta(2)}{}^{\dot\alpha} + \lambda \Phi^{\alpha\beta\dot\beta}
\Phi_{\beta\dot\beta}{}^{\dot\alpha} + 2 \Phi^{\alpha\dot\beta(2)}
\Phi^{\dot\alpha}{}_{\dot\beta(2)} ].
\end{eqnarray}
Now we consider a deformed Lagrangian (i.e. the sum of the free
Lagrangians where all curvatures are substituted by the covariant
ones)
\begin{equation}
\hat{\cal L} = - \frac{1}{12\lambda^3} \hat{\cal F}_{\alpha(3)}
\hat{\cal F}^{\alpha(3)} - \frac{1}{4\lambda}
\hat{\cal F}_{\alpha(2)\dot\alpha} \hat{\cal F}^{\alpha(2)\dot\alpha}
+ \frac{i}{4\lambda} \hat{R}_{\alpha(2)} \hat{R}^{\alpha(2)}
+ h.c. 
\end{equation}
and consider variation of this Lagrangian that do not vanish on-shell
\begin{equation}
\delta \hat{\cal F}^{\alpha(3)} = - R^{\alpha\beta} 
\rho^{\alpha(2)}{}_\beta, \qquad \delta \hat{R}^{\alpha(2)}
= 2b_1 {\cal F}^{\alpha\beta(2)} \rho^\alpha{}_{\beta(2)}.
\end{equation}
This produces
\begin{equation}
\delta \hat{\cal L} = [ \frac{1}{2\lambda^3} + \frac{2b_1}{\lambda^2}]
{\cal F}_{\alpha(2)\beta} R^{\beta\gamma} \rho^{\alpha(2)}{}_\gamma,
\end{equation}
so we have to put
$$
b_1 = - \frac{1}{4\lambda}.
$$
The deformed Lagrangian contains quadratic, cubic and quartic terms.
For the cubic part we obtain
\begin{eqnarray}
{\cal L}_1 &=& - \frac{1}{6\lambda^3} {\cal F}_{\alpha(3}
[ \omega^\alpha{}_\beta \Phi^{\alpha(2)\beta} + \lambda^2
h^\alpha{}_{\dot\alpha} \Phi^{\alpha(2)\dot\alpha}] \nonumber \\
 && - \frac{1}{2\lambda} {\cal F}_{\alpha(2)\dot\alpha}
[ \omega^\alpha{}_\beta \Phi^{\alpha\beta\dot\alpha}
+  \omega^{\dot\alpha}{}_{\dot\beta} \Phi^{\alpha(2)\dot\beta}
+ h_\beta{}^{\dot\alpha} \Phi^{\alpha(2)\beta} + \lambda
h^\alpha{}_{\dot\beta} \Phi^{\alpha\dot\alpha\dot\beta}] \nonumber \\
 && + \frac{b_1}{2\lambda^2} R_{\alpha(2)} [ 
\Phi^{\alpha\beta(2)} \Phi^\alpha{}_{\beta(2)} + 2\lambda^2
\Phi^{\alpha\beta\dot\alpha} \Phi^\alpha{}_{\beta\dot\alpha} +
\lambda^2 \Phi^{\alpha\dot\alpha(2)} \Phi^\alpha{}_{\dot\alpha(2)} ].
\end{eqnarray}
Note that we have terms with three derivatives, but they combine into
total derivative, as it is common in the massless case
\cite{Vas11,KhZ20a}. So the vertex has at most two derivatives, in
agreement with \cite{Met06a}. Using explicit expressions for
curvatures, integrating by parts, and using torsion zero condition, we
can bring this vertex to the following form:
\begin{eqnarray}
{\cal L}_1 &=& - \frac{1}{2\lambda} R^{\alpha\beta}
\Phi_{\alpha\dot\alpha(2)} \Phi_\beta{}^{\dot\alpha(2)} \nonumber \\
 && - D \Phi_{\alpha\beta\dot\alpha} h^\beta{}_{\dot\beta}
\Phi^{\beta\dot\alpha\dot\beta} - \frac{1}{2}
\Phi_{\alpha\dot\alpha\dot\beta} e_\alpha{}^{\dot\beta}
( \omega^\alpha{}_\beta \Phi^{\alpha\beta\dot\alpha} +
\omega^{\dot\alpha}{}_{\dot\gamma} \Phi^{\alpha(2)\dot\gamma})
\nonumber \\
 && + \lambda [ 3 \Phi_{\alpha\beta\dot\alpha}
e^\beta{}_{\dot\beta} h_\gamma{}^{\dot\beta}
\Phi^{\alpha\gamma\dot\alpha}- \Phi_{\alpha(2)\dot\alpha}
e_\beta{}^{\dot\alpha} h^\beta{}_{\dot\beta}
\Phi^{\alpha(2)\dot\beta}] + h.c. 
\end{eqnarray}
The terms in the last two lines correspond to the standard
covariantization of kinetic and mass-like terms in the free
Lagrangian:
\begin{equation}
e^{\alpha\dot\alpha} \Rightarrow e^{\alpha\dot\alpha} +
h^{\alpha\dot\alpha}, \qquad D \Rightarrow D + \omega^{\alpha(2)}
L_{\alpha(2)} + \omega^{\dot\alpha(2)} L_{\dot\alpha(2)},
\end{equation}
where $L_{\alpha(2)}$, $L_{\dot\alpha(2)}$ are generators of Lorentz
group. At the same time, the first line contains non-minimal
interactions that is necessary for the vertex to be gauge invariant
under the hypertransformations:
\begin{eqnarray}
\delta \Phi^{\alpha(2)\dot\alpha} &=& \omega^\alpha{}_\beta 
\rho^{\alpha\beta\dot\alpha} + \omega^{\dot\alpha}{}_{\dot\beta}
\rho^{\alpha(2)\dot\beta} + h_\beta{}^{\dot\alpha} 
\rho^{\alpha(2)\beta} + \lambda h^\alpha{}_{\dot\beta}
\rho^{\alpha\dot\alpha\dot\beta}, \nonumber \\
\delta h^{\alpha\dot\alpha} &=& b_1 [ - 2 \rho^{\alpha\beta(2)}
\Phi_{\beta(2)}{}^{\dot\alpha} + 2 \Phi^{\alpha\beta(2)}
\rho_{\beta(2)}{}^{\dot\alpha} + \lambda \Phi^{\alpha\beta\dot\beta}
\rho_{\beta\dot\beta}{}^{\dot\alpha} + h.c. ].
\end{eqnarray}
For the coupling constant of the non-minimal interactions we get
\begin{equation}
\kappa = - \frac{g}{2\lambda},
\end{equation}
where we restored the gravitational coupling constant $g$, which was
previously set to 1. Thus, by rescaling of the coupling constants, we
may obtain a non-singular flat limit, where only non-minimal
interactions survive. The same is true for the three massless fields
with arbitrary spins \cite{KhZ20a}.
 
\section{Partially massless case}

In this case one can construct two trivially gauge invariant vertices:
\begin{equation}
{\cal L} \sim h_1 E W^{\alpha(2)\beta(2)} {\cal F}_{\alpha(2)\gamma}
{\cal F}_{\beta(2)}{}^\gamma + h_2 R^{\alpha\beta}
 {\cal C}_{\alpha\gamma(2)} {\cal C}_\beta{}^{\gamma(2)} + h.c. 
\end{equation}
Using an explicit expression for curvature $R^{\alpha(2)}$,
integrating by parts and using differential identity for
${\cal C}^{\alpha(3)}$ one can show that the second vertex is
equivalent to the following combination of the abelian ones
\begin{equation}
{\cal L}_2 \sim - 2 \omega^{\alpha\beta} {\cal F}_{\alpha\gamma(2)}
{\cal C}_\beta{}^{\gamma(2)} + \lambda^2 e^{(\alpha}{}_{\dot\alpha}
h^{\beta)\dot\alpha} {\cal C}_{\alpha\gamma(2)} 
{\cal C}_\beta{}^{\gamma(2)} + h.c.
\end{equation}
It is easy to check that there are no other abelian vertices. 

Let us turn to the non-abelian vertices. For the cubic vertices in the
gauge invariant formalism, there also exist sufficient field
redefinitions to bring the vertex to abelian form \cite{Zin24a}. The
technical reason is that each gauge one-form has its own Stueckelberg
zero-form as it is natural when all gauge symmetries are spontaneously
broken. But in the partially massless case, some gauge one-forms do
not have their Stueckelberg zero-forms, and part of the gauge
symmetries remain unbroken. We investigated the most general
deformations of curvatures for both partially massless spin 5/2 as
well as massless spin 2, taking into account all possible field
redefinitions:
\begin{eqnarray}
\Phi^{\alpha(3)} &\Rightarrow& \Phi^{\alpha(3)} + \kappa_1
\phi^{\alpha(2)\beta} \omega^\alpha{}_\beta, \nonumber \\
\Phi^{\alpha(2)\dot\alpha} &\Rightarrow& \Phi^{\alpha(2)\dot\alpha}
+ \kappa_2 \phi^{\alpha(2)\beta} h_\beta{}^{\dot\alpha}, \\
\Phi^\alpha &\Rightarrow& \Phi^\alpha + \kappa_3 \phi^{\alpha\beta(2)}
\omega_{\beta(2)}, \nonumber 
\end{eqnarray}
\begin{eqnarray}
\omega^{\alpha(2)} &\Rightarrow& \omega^{\alpha(2)} + \kappa_4
\phi^\alpha{}_{\beta(2)} \Phi^{\alpha\beta(2)} + \kappa_5
\phi^{\alpha(2)}{}_\beta \Phi^\beta, \nonumber \\
h^{\alpha\dot\alpha} &\Rightarrow& h^{\alpha\dot\alpha} + \kappa_6
\phi^\alpha{}_{\beta(2)} \Phi^{\beta(2)\dot\alpha} + \kappa_7
e^{\alpha\dot\alpha} \phi_{\beta(3)} \phi^{\beta(3)}. 
\end{eqnarray}
It appears that there exists just one non-abelian deformation which
cannot be transformed into an abelian form. Not surprisingly, for the
partially massless spin 5/2 this deformation corresponds to the
standard minimal substitution rules.

The simplest way to construct such a vertex is to use our description
of the partially massless spin 5/2 a la Skvortsov-Vasiliev. It
contains only one-form fields, so the Fradkin-Vasiliev formalism
works without ambiguities, exactly as in the massless case. \\
{\bf Curvature deformations} For the partially massless spin 5/2 we
obtain (we again set coupling constant to be 1):
\begin{eqnarray}
\Delta {\cal F}^{\alpha(2)\dot\alpha} &=& \omega^\alpha{}_\beta
\Phi^{\alpha\beta\dot\alpha} + \omega^{\dot\alpha}{}_{\dot\beta}
\Phi^{\alpha(2)\dot\beta} + \frac{M}{2} h^\alpha{}_{\dot\beta}
\Phi^{\alpha\dot\alpha\dot\beta} + \frac{\tilde{m}}{3}
h^{\alpha\dot\alpha} \Phi^\alpha, \nonumber \\
\Delta {\cal F}^\alpha &=&  \omega^\alpha{}_\beta \Phi^\beta  
+ \tilde{m} h_{\beta\dot\alpha} \Phi^{\alpha\beta\dot\alpha} + 
\frac{3M}{2} h^\alpha{}_{\dot\alpha} \Phi^{\dot\alpha}. 
\end{eqnarray}
This gives for the hypertransformations
\begin{eqnarray}
\delta \Phi^{\alpha(2)\dot\alpha} &=& \omega^\alpha{}_\beta
\rho^{\alpha\beta\dot\alpha} + \omega^{\dot\alpha}{}_{\dot\beta}
\rho^{\alpha(2)\dot\beta} + \frac{M}{2} h^\alpha{}_{\dot\beta}
\rho^{\alpha\dot\alpha\dot\beta} + \frac{\tilde{m}}{3}
h^{\alpha\dot\alpha} \rho^\alpha, \nonumber \\
\delta \Phi^\alpha &=&  \omega^\alpha{}_\beta \rho^\beta  
+ \tilde{m} h_{\beta\dot\alpha} \rho^{\alpha\beta\dot\alpha} + 
\frac{3M}{2} h^\alpha{}_{\dot\alpha} \rho^{\dot\alpha}.
\end{eqnarray}
For the massless spin 2 the solution is also unique:
\begin{eqnarray}
\Delta R^{\alpha(2)} &=& b_1 \Phi^{\alpha\beta\dot\alpha}
\Phi^\alpha{}_{\beta\dot\alpha} + \frac{b_1}{2} 
\Phi^{\alpha\dot\alpha(2)} \Phi^\alpha{}_{\dot\alpha(2)} - 
\frac{b_1}{3} \Phi^\alpha \Phi^\alpha, \nonumber \\
\Delta T^{\alpha\dot\alpha} &=& \frac{b_1}{M}
\Phi^{\alpha\beta\dot\beta} \Phi_{\beta\dot\beta}{}^{\dot\alpha}
- \frac{5b_1}{2\tilde{m}} (\Phi^{\alpha\beta\dot\alpha} \Phi_\beta +  
\Phi^{\alpha\dot\alpha\dot\beta} \Phi_{\dot\beta}) - \frac{b_1}{M}
\Phi^\alpha \Phi^{\dot\alpha}. 
\end{eqnarray}
{\bf Invariance of the deformed Lagrangian} Let us consider a deformed
Lagrangian:
\begin{equation}
\hat{\cal L} = - \frac{1}{2M} \hat{\cal F}_{\alpha(2)\dot\alpha}
\hat{\cal F}^{\alpha(2)\dot\alpha} + \frac{1}{3M} \hat{\cal F}_\alpha
\hat{\cal F}^\alpha + \frac{i}{4\lambda^2} \hat{R}_{\alpha(2)}
\hat{R}^{\alpha(2)} + h.c. 
\end{equation}
Non-trivial hypertransformations for curvatures (taking into account
torsion zero condition) are
\begin{eqnarray}
\delta \hat{\cal F}^{\alpha(2)\dot\alpha} &=& R^\alpha{}_\beta
\rho^{\alpha\beta\dot\alpha} + R^{\dot\alpha}{}_{\dot\beta}
\rho^{\alpha(2)\dot\beta}, \qquad \delta \hat{\cal F}^\alpha =
R^\alpha{}_\beta \rho^\beta, \nonumber \\
\delta \hat{R}^{\alpha(2)} &=& 2b_1 {\cal F}^{\alpha\beta\dot\alpha}
\rho^\alpha{}_{\beta\dot\alpha} - \frac{2b_1}{3} {\cal F}^\alpha
\rho^\alpha.
\end{eqnarray}
They produce the following variations:
\begin{equation}
\delta \hat{\cal L} = [ \frac{b_1}{\lambda^2} + \frac{1}{M}]
R_{\alpha(2)} {\cal F}^{\alpha\beta\dot\alpha}
\rho^\alpha{}_{\beta\dot\alpha} - [\frac{b_1}{3\lambda^2}
+ \frac{1}{3M}] R_{\alpha2)} {\cal F}^\alpha \rho^\alpha + h.c.
\end{equation}
Thus we set (recall that $M^2 = \lambda^2)$
\begin{equation}
b_1 = - M.
\end{equation}
{\bf Cubic vertex} For the cubic part of the deformed Lagrangian we
obtain\begin{eqnarray}
{\cal L}_1 &=& - \frac{1}{M} {\cal F}_{\alpha(2)\dot\alpha}
[ \omega^\alpha{}_\beta \Phi^{\alpha\beta\dot\alpha} +
 \frac{M}{2} h^\alpha{}_{\dot\beta} \Phi^{\alpha\dot\alpha\dot\beta}
+ \frac{\tilde{m}}{3} h^{\alpha\dot\alpha} \Phi^\alpha ]
 + \frac{1}{M} {\cal F}_{\alpha\dot\alpha(2)} \omega^\alpha{}_\beta
\Phi^{\beta\dot\alpha(2)} \nonumber \\
 && + \frac{2}{3M} {\cal F}_\alpha [ \omega^\alpha{}_\beta
\Phi^\beta + \tilde{m} h_{\beta\dot\alpha}
\Phi^{\alpha\beta\dot\alpha} + \frac{3M}{2} h^\alpha{}_{\dot\alpha}
\Phi^{\dot\alpha}] \nonumber \\
 && + \frac{b_1}{2\lambda^2} R_{\alpha(2)} 
[ \Phi^{\alpha\beta\dot\alpha} \Phi^\alpha{}_{\beta\dot\alpha}
+ \frac{1}{2} \Phi^{\alpha\dot\alpha(2)} \Phi^\alpha{}_{\dot\alpha(2)}
- \frac{1}{3} \Phi^\alpha \Phi^\alpha ] + h.c.
\end{eqnarray}
Note that contrary to the massless case there are no terms with more
than two derivatives. Now using explicit expressions for the
curvatures, integrating by parts and using torsion zero condition one
can bring this vertex into the suggestive form:
\begin{eqnarray}
 {\cal L}_1 &=& - \frac{1}{M} \Phi_{\alpha\dot\alpha(2)}
R^{\alpha\beta} \Phi_\beta{}^{\dot\alpha(2)} \nonumber \\
 && - D \Phi_{\alpha\beta\dot\alpha} h^\alpha{}_{\dot\beta}
\Phi^{\beta\dot\alpha\dot\beta} + D \Phi_\alpha 
h^\alpha{}_{\dot\alpha} \Phi^{\dot\alpha} \nonumber \\
 && + \Phi_{\alpha\beta\dot\alpha} e^\beta{}_{\dot\beta}
\omega^\alpha{}_\gamma \Phi^{\gamma\dot\alpha\dot\beta}
+ \Phi_{\alpha\beta\dot\alpha} e^\beta{}_{\dot\beta}
\omega^{(\dot\alpha}{}_{\dot\gamma} \Phi^{\alpha|\dot\beta)\dot\gamma}
- \Phi_\alpha e^\alpha{}_{\dot\alpha} 
\omega^{\dot\alpha}{}_{\dot\beta} \Phi^{\dot\beta} \nonumber \\
 && - \frac{3M}{4} \Phi_{\alpha\beta\dot\alpha} 
e^{(\alpha}{}_{\dot\beta} h^{\gamma)\dot\beta} 
\Phi_\gamma{}^{\beta\dot\alpha} + \frac{M}{4} 
\Phi_{\alpha(2)\dot\alpha}  e_\beta{}^{(\dot\alpha} 
h^{\beta|\dot\beta)} \Phi^{\alpha(2)}{}_{\dot\beta} \nonumber \\
 && + 2\tilde{m} \Phi_{\alpha\beta\dot\beta} 
e^\alpha{}_{\dot\alpha} h^{\beta\dot\alpha} \Phi^{\dot\beta}
 + 3M \Phi_\alpha e^\alpha{}_{\dot\alpha}
h^{\beta\dot\alpha} \Phi_\beta + h.c. 
\end{eqnarray}
Indeed, in the first line we see the non-minimal interaction, while
all other terms correspond to the standard covariantization for the
kinetic and mass-like terms in the free Lagrangian.

\section{Massive case}

For the cubic vertices with one massless and two massive fields we
always have sufficient field redefinitions to bring the vertex to
abelian form. We begin our analysis with the  trivially gauge
invariant ones. They can be written in terms of the gauge invariant
zero-forms:
\begin{eqnarray*}
{\cal L} &\sim& E W^{\alpha(2)\beta(2)} [ h_1 Y_{\alpha(2)\gamma(3)}
Y_{\beta(2)}{}^{\gamma(3)} + h_2 Y_{\alpha(2)\gamma(2)\dot\alpha}
Y_{\beta(2)}{}^{\gamma(2)\dot\alpha} \\
 && \qquad + h_3  Y_{\alpha(2)\gamma\dot\alpha(2)} 
Y_{\beta(2)}{}^{\gamma\dot\alpha(2)} + h_4 Y_{\alpha(2)\dot\alpha(3)}
Y_{\beta(2)}{}^{\dot\alpha(3)}] + h.c.
\end{eqnarray*}
but we prefer their on-shell equivalent version:
\begin{equation}
{\cal L} \sim h_1 W^{\alpha(2)\beta(2)} {\cal F}_{\alpha(2)\gamma}
{\cal F}_{\beta(2)}{}^\gamma + {\cal R}^{\alpha\beta} [ h_2
{\cal C}_{\alpha\gamma(2)} {\cal C}_\beta{}^{\gamma(2)} + h_3
{\cal C}_{\alpha\gamma\dot\alpha} {\cal C}_\beta{}^{\gamma\dot\alpha}
+ h_4 {\cal C}_{\alpha\dot\alpha(2)} {\cal C}_\beta{}^{\dot\alpha(2)}]
+ h.c. 
\end{equation}
Then using explicit expression for ${\cal R}^{\alpha(2)}$,
integrating by parts and using differential identities for curvatures
one can show that the three last vertices are equivalent to some
combinations of the abelian ones:
\begin{eqnarray}
{\cal L}_2  &\approx& 2 \Omega^{\alpha\beta} 
[ - {\cal F}_{\alpha\gamma(2)} {\cal C}_\beta{}^{\gamma(2)} +
\frac{3a_0}{16} e_\alpha{}^{\dot\alpha} {\cal C}_{\gamma(2)\dot\alpha}
{\cal C}_\beta{}^{\gamma(2)} ] + \lambda^2 e^{(\alpha}{}_{\dot\alpha}
H^{\beta)\dot\alpha} {\cal C}_{\alpha\gamma(2)} 
{\cal C}_\beta{}^{\gamma(2)} + h.c., \nonumber \\
{\cal L}_3 &\approx& 2 \Omega^{\alpha\beta} 
[ - e_\alpha{}^{\dot\alpha}  {\cal C}_{\gamma(2)\dot\alpha} 
{\cal C}_\beta{}^{\gamma(2)} + M e_\alpha{}^{\dot\alpha} 
{\cal C}_{\gamma\dot\alpha\dot\beta} 
{\cal C}_\beta{}^{\gamma\dot\beta} ] + \lambda^2 
e^{(\alpha}{}_{\dot\beta} H^{\beta)\dot\beta} 
{\cal C}_{\alpha\gamma\dot\alpha} {\cal C}_\beta{}^{\gamma\dot\alpha}
 + h.c., \\
{\cal L}_4  &\approx& 2 \Omega^{\alpha\beta} [ e_\alpha{}^{\dot\alpha}
{\cal C}_{\dot\alpha\dot\beta(2)} {\cal C}_\beta{}^{\dot\beta(2)}
- M e_\alpha{}^{\dot\alpha} {\cal C}_{\gamma\dot\alpha\dot\beta}
{\cal C}_\beta{}^{\gamma\dot\beta} ] + \lambda^2 
e^{(\alpha}{}_{\dot\beta} H^{\beta)\dot\beta} 
{\cal C}_{\alpha\dot\alpha(2)} {\cal C}_\beta{}^{\dot\alpha(2)} + h.c.
\nonumber
\end{eqnarray}
\begin{eqnarray}
{\cal L}_a &=& \Omega^{\alpha\beta} [ g_1 {\cal F}_{\alpha\gamma(2)}
{\cal C}_\beta{}^{\gamma(2)} + g_2 e_\alpha{}^{\dot\alpha}
{\cal C}_{\gamma(2)\dot\alpha} {\cal C}_\beta{}^{\gamma(2)}
+ g_3 e_\alpha{}^{\dot\alpha} {\cal C}_{\gamma\dot\alpha\dot\beta}
{\cal C}_\beta{}^{\gamma\dot\beta} + g_4 e_\alpha{}^{\dot\alpha}
{\cal C}_{\dot\alpha\dot\beta(2)} {\cal C}_\beta{}^{\dot\beta(2)}]
\nonumber \\
 && + f_1 H^{\alpha\dot\alpha} {\cal F}_{\alpha\beta(2)}
{\cal C}^{\beta(2)}{}_{\dot\alpha} 
+ e^{(\alpha}{}_{\dot\alpha} H^{\beta)\dot\alpha} [ f_2
{\cal C}_{\alpha\gamma(2)} {\cal C}_\beta{}^{\gamma(2)} + f_3
{\cal C}_{\alpha\gamma\dot\beta} {\cal C}_\beta{}^{\gamma\dot\beta} +
f_4 {\cal C}_{\alpha\dot\beta(2)} {\cal C}_\beta{}^{\dot\beta(2)}]
\end{eqnarray}
The requirement that this expression be gauge invariant leads to four
solutions three of which being equivalent to the trivially gauge
invariant ones. Thus we have five possible cubic vertices.

\noindent
{\bf Minimal vertex} To construct a minimal vertex (i.e. a vertex
which contains standard minimal interactions and has the minimum
number of derivatives), we use the fact that the standard substitution
rules work completely non-linearly, not just at the cubic level. Thus
we take the Lagrangian (\ref{lag_m}), the gauge transformations
(\ref{gauge_1}), (\ref{gauge_2}),  and the curvatures (\ref{cur_1}),
(\ref{cur_2}), where the frame $e^{\alpha\dot\alpha}$ and Lorentz
covariant derivative are now dynamical. We still assume that torsion
is zero, so that the only source of non-invariance of the Lagrangian
is the non-commutativity of covariant derivatives. Then by
straightforward calculations we obtain:
\begin{equation}
\delta {\cal L}_0 =  2 e_\beta{}^{\dot\beta}
 (R^\alpha{}_\gamma \Phi_{\gamma\dot\alpha\dot\beta}
 + R_{\dot\alpha}{}^{\dot\gamma} \Phi_{\alpha\dot\beta\dot\gamma}
+ R_{\dot\beta}{}^{\dot\gamma} \Phi_{\alpha\dot\alpha\dot\gamma})
\rho^{\alpha\beta\dot\alpha}  - 2 e_\alpha{}^{\dot\alpha}
R_{\dot\alpha}{}^{\dot\beta} \Phi_{\dot\beta} \rho^\alpha + h.c.
\end{equation}
Using our experience from the partially massless case we add the
following ansatz for the non-minimal interactions:
\begin{equation}
{\cal L}_1 = \kappa R_{\alpha\beta} \Phi^{\alpha\dot\alpha(2)}
\Phi^\beta{}_{\dot\alpha(2)} + h.c.
\end{equation}
Calculating its variations we obtain
\begin{eqnarray}
\delta {\cal L}_1 &=& 2\kappa R_{\alpha\beta} 
\Phi^\alpha{}_{\dot\alpha(2)} [ D \rho^{\beta\dot\alpha(2)} +
e^\beta{}_{\dot\beta} \rho^{\dot\alpha(2)\dot\beta} + \frac{M}{2}
e_\gamma{}^{\dot\alpha} \rho^{\beta\gamma\dot\alpha} +
\frac{\tilde{m}}{3} e^{\beta\dot\alpha} \rho^{\dot\alpha}] + h.c.
\nonumber \\
 &\approx&  2\kappa R_{\dot\alpha\dot\beta} [ e_\beta{}^{\dot\alpha}
\Phi^{\alpha(2)\beta} + \frac{M}{2} e^\alpha{}_{\dot\gamma}
\Phi^{\alpha\dot\alpha\dot\gamma} + \frac{\tilde{m}}{3}
e^{\alpha\dot\alpha} \Phi^\alpha] \rho_{\alpha(2)}{}^{\dot\beta}
\nonumber \\
 && + M\kappa R_{\alpha\beta} \Phi^\alpha{}_{\dot\alpha(2)}
e_\gamma{}^{\dot\alpha} \rho^{\beta\gamma\dot\alpha} - 2\kappa
R_{\dot\alpha\dot\beta} \Phi_{\alpha(2)}{}^{\dot\alpha}
[ e_\beta{}^{\dot\beta} \rho^{\alpha(2)\beta} + \frac{\tilde{m}}{3}
e^{\alpha\dot\beta} \rho^\alpha ] + h.c.
\end{eqnarray}
Recall that $\approx$ means ''up to total derivatives and terms which
vanish on-shell''. Then if we set
\begin{equation}
\kappa = - \frac{1}{M},
\end{equation}
we get
\begin{eqnarray}
\delta ({\cal L}_0 + {\cal L}_1) &=& 
e^\gamma{}_{\dot\alpha} R_{\alpha\gamma} 
[ - \frac{2}{M} \rho^{\alpha\beta(2)} \Phi_{\beta(2)}{}^{\dot\alpha}
+ \frac{2}{M} \Phi^{\alpha\beta(2)} \rho_{\beta(2)}{}^{\dot\alpha} 
- 2 \rho^{\alpha\beta\dot\beta} \Phi_{\beta\dot\beta}{}^{\dot\alpha} 
\nonumber \\
 && \qquad + \frac{4\tilde{m}}{3M} (\Phi^{\alpha\beta\dot\alpha}
\rho_\beta - \rho^{\alpha\beta\dot\alpha} \Phi_\beta) - 2
\Phi^{\dot\alpha} \rho^\alpha ] + h.c.
\end{eqnarray}
Now all these terms can be compensated by the following corrections
\begin{eqnarray}
\delta H^{\alpha\dot\alpha} &=&  
[  \frac{1}{M} \rho^{\alpha\beta(2)} \Phi_{\beta(2)}{}^{\dot\alpha}
- \frac{1}{M} \Phi^{\alpha\beta(2)} \rho_{\beta(2)}{}^{\dot\alpha} 
+ \rho^{\alpha\beta\dot\beta} \Phi_{\beta\dot\beta}{}^{\dot\alpha} 
\nonumber \\
 && \qquad - \frac{2\tilde{m}}{3M} (\Phi^{\alpha\beta\dot\alpha}
\rho_\beta - \rho^{\alpha\beta\dot\alpha} \Phi_\beta) +
\Phi^{\dot\alpha} \rho^\alpha ] + h.c.
\end{eqnarray}
Note that in the framework of the Fradkin-Vasiliev formalism these
corrections correspond to the following deformation for the torsion:
\begin{equation}
\Delta T^{\alpha\dot\alpha} =  \frac{1}{M} \Phi^{\alpha\beta(2)}
\Phi_{\beta(2)}{}^{\dot\alpha} + \Phi^{\alpha\beta\dot\beta}
\Phi_{\beta\dot\beta}{}^{\dot\alpha} + \frac{2\tilde{m}}{3M}
\Phi^{\alpha\beta\dot\alpha} \Phi_\beta + \Phi^{\dot\alpha}
\Phi^\alpha + \dots + h.c. 
\end{equation}
where dots stand for the contributions of the zero-forms.

For the coefficient at the non-minimal interactions we get
\begin{equation}
\kappa = - \frac{g}{\sqrt{m^2 - 4\Lambda}},
\end{equation}
where we restore the gravitational coupling constant, which was
previously set to 1. Thus, for a non-zero cosmological constant
$\Lambda$, we have a smooth massless limit (unitary in $AdS_4$), while
for non-zero mass we have a smooth flat limit. The special point 
$m^2 = 4\Lambda$ corresponds to the boundary of the unitarity region.
As in the massless case, the only way to get a non-trivial result is
by rescaling of the coupling constant which leaves us with the 
non-minimal interactions only.

\section{Conclusion}

In this work, we investigated the gravitational interactions for
massive spin 5/2, including partially massless and massless limits. We
used the gauge invariant frame-like formalism for the description of
massive spin 5/2 and the Fradkin-Vasiliev formalism to construct
interactions. One of the technical problems in using a gauge
invariant description of massive fields, is related to field
redefinitions containing Stueckelberg zero-forms. In particular, for
any vertex with one massless and two massive fields (as it is the case
for gravitational interactions) one always has enough field
redefinitions to bring the vertex into an abelian form. We used this
fact for the classification of possible vertices. On the other
hand, to construct a minimal vertex (i.e. containing standard minimal
gravitational interactions and non-minimal ones with a minimum
number of derivatives) we used a fairly common down-up approach. And
it turned out that the simplest and most straightforward way to find
the appropriate form of the non-minimal interactions was to consider
the partially massless case within the Skvortsov-Vasiliev version of
the frame-like formalism. Our results indicate that there is a
specific point $m^2 = 4\Lambda$, which corresponds to the boundary of
unitary allowed region, where standard minimal interactions are
absent, leaving only non-minimal ones. In the companion paper, we will
apply the same strategy to the massive spin 3 field, including
its partially massless and massless limits.

\end{document}